**Deep Learning Body Region Classification of MRI and CT examinations**

Philippe Raffy[1], PhD, Jean-François Pambrun[1], PhD, Ashish Kumar[1], David Dubois[1], PhD, Jay Waldron Patti[2], MD, Robyn Alexandra Cairns[3], MD, Ryan Young[1]

[1] Change Healthcare, Enterprise Imaging Solutions, Vancouver, Canada

[2] Mecklenburg Radiology Associates, Charlotte, NC

[3] University of British Columbia, Vancouver, British Columbia, Canada

| | |
|---|---|
| **Corresponding Author:** | David Dubois, PhD |
| | Change Healthcare, Enterprise Imaging Solutions |
| | 10711 Cambie Road |
| | Richmond, BC V6X 3G5 Canada |
| | Ph.: +1 612 670 3311 |
| | Email: David.Dubois1@changehealthcare.com |

**Manuscript type**: Original research

**Word count for Text** (Introduction to Discussion): 2,946 words

**Unblinded acknowledgments**: The authors thank Clementine Raffy for her image labeling support.

**Key points:**

- An off the shelf deep learning model can achieve statistically meaningful anatomic classification results of over 90% sensitivity on CT and MRI sets completely disjoint of training sets.
- Generalization is facilitated by using images sets with diverse acquisition protocols and manufacturers.
- Upper and lower extremity images have similarities that confuse model yet results for these regions validate its effectiveness.

**Abstract** (245 words)

Purpose: Standardized body region labelling of individual images provides data that can improve human and computer use of medical images. A CNN-based classifier was developed to identify body regions in CT and MRI.

Material and Methods: 17 CT (18 MRI) body regions covering the entire human body were defined for the classification task. Three retrospective databases were built for the AI model training, validation, and testing, with a balanced distribution of studies per body region. The test databases originated from a different healthcare network. Sensitivity and specificity of the classifier was evaluated for patient age, patient gender, institution, scanner manufacturer, contrast, slice thickness, MRI sequence, and CT kernel.

Results: The data included a retrospective cohort of 2,891 anonymized CT cases (training: 1,804 studies, validation: 602 studies, test: 485 studies) and 3,339 anonymized MRI cases (training: 1,911 studies, validation: 636 studies, test: 792 studies). 27 institutions from primary care hospitals, community hospitals and imaging centers contributed to the test datasets. The data included cases of all genders in equal proportions and subjects aged from 18 years old to +90 years old. Image-level weighted sensitivity of 92.5% (92.1 – 92.8) for CT, and 92.3% (92.0 – 92.5) for MRI, and weighted specificity of 99.4% (99.4 – 99.5) for CT and 99.2% (99.1 – 99.2) for MRI were achieved. The classification results were robust across all body regions and confounding factors.

Conclusion: Deep learning models can classify CT and MRI images by body region including lower and upper extremities with high accuracy.

## INTRODUCTION (400 words)

Improved accuracy of anatomic classification of imaging data is required to improve the usability of medical imaging data throughout imaging and enterprise workflows (1). Body region categorization is foundational information in radiology workflows such as ordering, triage, acquisition, interpretation and report communication. Anatomical metadata is used by humans and machines for selection of relevant prior studies, organizing display protocols, registration and navigation of studies, and contextual interpretation and reporting. It is also a pre-requisite for diagnostic AI such as Computer Aided Detection (2) and Computer Aided Triage systems (3) that work for specific anatomies and clinical indications

However, in practice, body region metadata documented in DICOM labelling of series and studies is often inconsistent within and across institutions. For example, studies for knee imaging may be labelled as "knee MRI" in one institution and "lower extremity CT" in another institution. Similarly, a "CT elbow" may be called "CT upper extremity" in another institution. As important, the anatomical content of an imaging study cannot always be summarized to one body region. For example, an abdomen exam may contain images of the chest that may be important to review if a new lung nodule is found on subsequent chest imaging. Accurate labelling of the anatomy covered by a study could therefore have an impact on the management of incidental findings. In short, current inconsistencies in the way anatomical metadata is documented makes the search and access to anatomically matching studies cumbersome and inefficient and has a detrimental impact on a radiologist's accuracy and productivity.

To remedy this, we propose to develop a pixel-based AI to automatically identify body regions in CT and MRI studies. Previous attempts to tackle this classification problem using supervised and unsupervised deep learning techniques have shown accuracy results ranging from 72% to 92% in limited settings (4, 5, 6, 7, 8). Limitations include modality (CT), study design (lack of independent test database, repeat studies, lack of information about image inclusion/exclusion criteria), neural network architecture (legacy neural nets), the size of the database (<1,700 total patients, ≤100,000 total images), the number of body region classes (≤12, no upper extremities), and the extent of clinical protocols covered (mostly thin slice CT acquisitions, no contrast medium). The main study's objective is to demonstrate that the anatomical classifier can reach state-of-the-art accuracy or weighted sensitivity greater than 90% across a large spectrum of body regions, patient demographics, patient comorbidities, and clinical protocols.

**MATERIALS AND METHODS** (1287 words)

**Study Design**

The performance of AI models was evaluated in a retrospective standalone study using manually defined ground truth data. Three databases were collected for the AI model training and evaluation: training, validation, test (holdout set). The selection of body regions was established with the goal of covering the entire human body. 17 CT (18 MRI) body regions were considered for the classification task: abdomen, breast (MRI only), calf/lower leg, chest, elbow, foot, forearm, hand, head, humerus/arm, knee, neck, pelvis/hip, shoulder, spine cervical, spine thoracic, spine lumbar, thigh/upper leg. Non-contrast scans were collected for all body regions and complemented with MRA and CTA contrast datasets for head, neck, chest, and abdomen.

**Data**

The data used for training and validation originated from a large cohort of 63,699 de-identified studies (CT: 28,211 patients; MRI: 35,488 patients) from one healthcare system and its affiliates (University of Wisconsin Health). Patients underwent contrast or non-contrast imaging between 1997 and 2020 (median: 2017) and came from a tertiary care hospital as well as a smaller affiliated hospital and several outpatient imaging centers. A well-controlled selection process was used to build the training and validation datasets. In the first phase, the data was randomly selected from this large pool to ensure a balanced distribution of body regions. In the second phase, several iterations of active learning (9, 10) were applied to the large cohort of unlabeled data until the accuracy goal was reached. This step was added to enrich the datasets with more complex anatomical cases.

The test databases were collected from a different source, e.g., United Point Health system (UPH) and its affiliates, so as to address the general issue of model under specification (11). Because of the over-representation of datasets acquired with a GE scanner in the training and validation datasets, the instruction was to focus on collecting datasets from multiple institutions where a GE scanner was not predominantly used. The same attention to collecting a balanced representation of all the body parts was given. The CT and MRI imaging data from 3,003 patients scanned between 2016 and 2020

(median: 2020) was collected and served as a pool of data to build the test databases. These patients came from primary care hospitals, critical care hospitals, and imaging centers.

This all-comers study was designed with the intent to be as inclusive as possible and clinically relevant. All randomly selected patients were included in the study irrespective of their demographic (ex: ethnicity) or comorbidities. Due to the under-representation of pediatric cases, test databases were only composed of adult patients. The same inclusive approach was followed for acquisition protocols, orientations (supine, prone, lateral), CT kernels and MRI sequences. The only exclusions were applied to protocols where the anatomical region was faded and challenging to distinguish such as flow protocols and MRI elastography (Supplemental Materials – Inclusion and exclusion criteria). Models were trained on 2D transversal slices. Axial images with an angle up to 45 degrees were included to cover oblique acquisitions. Multiplanar reformat series such as coronal and sagittal series as well as series used to aid planning of the acquisition such as scout, calibration, and quality control series were excluded. Similarly, all post-processed series such as perfusion maps, reformat, 3D reconstruction, secondary capture, CAD, CINE and movies were not considered. Image exclusion criteria were image stored in a format 8 bits or lower, image with multiple channels (RGB or other not grayscale), no pixel data, very limited amount of pixel data (<1,000 pixels), and no compatible codec (anything not JPEG lossless, Raw RLE). All data were de-identified by partners on site prior to any processing to comply with HIPAA requirements. The de-identification schema followed the one used in the Cancer Imaging Archive initiative (12).

**Ground Truth**

The ground truth was labeled based on clearly defined anatomical landmarks (Table 1) using a home-grown annotation software (Fig 1). To avoid the effect of inter-reader variability, all the datasets were labeled by one image annotator with a long-time experience of developing CAD solutions (PR). Pediatric patients under 10 years old were reviewed and edited independently by an experienced pediatric radiologist (RAC) but were left out of the final analysis because of insufficient number of samples. To evaluate the accuracy and consistency of the ground truth, an independent review of a random subset of the labeled dataset was conducted by an experienced radiologist (JWP). Consistent with peer review practices (13), 2.5% of the labeled data was reviewed with an equal number of studies assigned for all body regions.

**Data Partitions**

Repeat studies in the validation and test databases were removed so one study corresponds to one patient. Labeled datasets were organized according to the main body region and sorted according to study size. A 75/25 split between the training and validation patient datasets was performed for each body region. To estimate the size of the test databases, a strict survey study sampling model (14) was used with the assumptions of a model at least 90% accurate and a 95% confidence interval with a 10% relative error. Based on this model it was determined that at least 7,600 images per body region for CT, and at least 3,600 images per body region for MRI were needed.

**Model**

The classification task is composed of multiple stages that are detailed in Fig 2. As a first step, we used the standard ResNet50V2 model (15) in a multi-class framework. Following the 2D CNN classifier, a few post-processing steps at the series level were applied. First, a rule engine merged the abdomen-chest class to either the abdomen or chest class and classified an entire series as breast if at least 50% of the images in the series were classified as such. Second, a routine was applied to remove MRI series with uncertain results based on the evaluation of decision margin. The goal was to discard series with little anatomical content and severe artifacts. Last, outlier labels not related to the anatomical content of the series were detected and the image predictions smoothed out so a continuity in classified labels could be observed throughout the series.

**Training**

We enriched our dataset by applying spatial deformations to a random set of images in each training epoch. These transforms include rotation within an angle of $\pm \pi/10$, translation and shear with a maximum of 10% in image size in both directions, scaling with a maximum of 20% in image size in both directions, and bilinear interpolation. The transformations were applied using built in TensorFlow library functions. We used the transfer learning approach and model weights from pretrained Resnet50V2 model developed for vision benchmark ImageNet dataset. The training hyperparameters are listed in Supplemental Materials – Training Parameters. The loss function used is categorical cross entropy which is well-suited for the multi-class case.

**Evaluation**

We applied the models to the test datasets and evaluated them by computing the average weighted values of the following performance metrics: sensitivity and specificity. Because of the data distribution, accuracy and aggregated weighted sensitivity are the same so only the latter is reported in the paper. Results were derived by body region, institution, patient demographics, and acquisition parameters (manufacturer, contrast, CT kernel, slice thickness, sequence type). Performance metrics and their corresponding confidence intervals were determined using the jackknife bootstrap resampling method (16). Random image sampling every 10mm was performed at the series level to reduce the impact of strongly correlated images and provide more realistic statistical results. This is consistent with the 7.5mm sampling approach reported in (17). In order to reduce the inter-series correlation, only one of the series (randomly selected at each sampling iteration) in the subsampled dataset was kept. The correlation and correlation significance between the model's accuracy and each confounding factor was assessed using Cramer's V and Pearson's chi-squared statistical test.

## RESULTS (576 words)

**Data**

The data consisted of 2,891 CT cases (training: 1,804 studies, validation: 602 studies, test: 485 studies) and 3,339 MRI cases (training: 1,911 studies, validation: 636 studies, test: 792 studies). Flowcharts in the Supplemental Materials (Inclusion and Exclusion Criteria) show the distribution of images after the different stages of series and image exclusion criteria. The evaluation of the ground truth revealed a total of 4 labeling errors out of 1,455 CT and MRI labeled studies, which represents an error rate of 0.3% per study. There were no cases where the labeling discrepancy would be considered clinically relevant, if for example, two physicians were discussing a patient and using the discrepant labels interchangeably.

Distributions of images and results by confounding factors for the Test sets can be found in Tables 2-3. 27 institutions contributed to each CT and MRI Test dataset. For CT, 56% of datasets came from primary care hospitals and 44% from critical access hospitals and imaging centers while for MRI, 55% of the datasets came from primary care hospitals. Gender parity was respected for the CT database. A slight over-representation of female gender was noticed for the MRI database (56.1%). The age coverage ranged from 18 years old to +90 years, roughly following the distribution of imaging tests in US Healthcare Systems (18). Compared to the development databases (Supplemental Materials – Distribution Development Databases), the test databases differed in some key areas. For CT, acquisitions mostly originated from Siemens and non-GE scanners (87.6%, +80.1%) with a larger proportion of older adults ≥65 years (45.6%, +9.3%), intermediate slice thickness (2mm < slice thickness < 5mm) (54.8%, +46.3%), and non-contrast imaging (76.5%, +9.1%). For MRI, acquisitions mostly originated from Siemens and non-GE scanner (83.5%, +69%), a larger proportion of older adults (31.7%, +11.7%), cases with intermediate slice thickness (68.5%, +20.3%), and non-contrast imaging (84.0%, +6.4%).

**Model Performance**

An overall body region image-level sensitivity of 92.5% (92.1 – 92.8) was achieved for CT and 92.3% (92.0 – 95.6) for MRI. The post-processing stages contributed to about 1.1% (CT) to 1.6% (MRI) improvement in classification accuracy. Classification results by body region are listed in Table 4. Accuracy results were consistent across body regions. Head and breast images have very discernable features, so they tend to be more accurately classified than other body regions such as the neck and extremities. Confidence intervals for sensitivity was lower (upper bound did not include 90%) for the following anatomical regions: CT cervical spine, CT forearm, CT pelvis, MRI cervical spine, MRI forearm and MRI neck.

No formal association was found between classification accuracy and CT institution, CT kernel, MRI contrast. However, statistically superior classification results were noticed in a few instances with Cramer's V correlation ranging from negligible (V<0.05) to moderate (V=0.17). For CT, that was the case for datasets with older (>= 65) age (p<0.001, V=0.041) with contrast (p<0.001, V=0.042), and thick (>= 5mm) slice (p=<0.001, V=0.048). For MRI, imaging centers (p<0.001, V=0.064), 44 years and older (p<0.001, V=0.087), Philips manufacturer (p=0.001, V=0.076), thin slices (p<0.001, V=0.0838) and Inversion and MRA sequences (p<0.001, V=0.179) exhibited better classification performance. For some of the classes in the test sets, the association between accuracy and factors such as manufacturer and MRI sequence could not be reliably assessed: Hitachi and Canon scanner manufacturers and In and Out of Phase MRI sequences. In spite of these limitations, the evaluation of accuracy results and confidence intervals points to performance robustness across age, manufacturer, CT slice thickness and MRI sequence categories.

**DISCUSSION** (683 words)

This work demonstrates how a deep learning CNN based classifier can achieve overall state-of-the-art accuracy greater than 90% in identifying body regions in CT and MRI studies while covering the entire human body and a large spectrum of acquisition protocols across separate institutions. This is the first known attempt to a) provide a solution for MRI, and b) offer a clinically relevant solution that covers numerous classes, in particular extremities that are more prone to misclassifications.

There are several reasons why extremities are challenging to classify. First, extremities can be imaged in multiple orientations depending on the patient's clinical condition. This means that in some cases, an axial series may present in the form of an oblique axial or even another MPR view. The training database was built to account for such a diversity of orientations. Second, extremities are subject to fractures, surgical implants and imputation, which can be sources of body region misclassification. Third, images of upper and lower extremities have structural similarities that can make them look similar in some circumstances. This is the case for joints, paired long bones of the calf and forearm, single long bones of the thigh and arm, and knee and elbow. Fourth, the boundaries between body regions for the upper and lower extremities are less distinct than between other body parts.

Although slice thickness and FOV can be informative, the differentiation between a neck and a cervical spine study can be sometimes difficult in particular for soft tissue reconstructions. The clinical impact may be minimal though since both body regions represent the same anatomical region. At the time of field deployment, the user can decide to configure the application so that these two classes are merged.

The study has some limitations. First, the multiclass framework is, by design, not well suited to identify multiple regions in an image. This is a limitation when dealing with whole body studies. Second, the ML models were trained and evaluated on axial images. The same way data from the axial plane are reconstructed to non-axial planes, anatomical predictions determined in the axial plane can be easily ported to coronal and sagittal planes. Third, there are classes in the databases that are under-represented such as Hitachi and Canon scanner manufacturers, CT thick slices ≥5 mm, and In and Out of Phase MRI sequences. Additional data is needed to complement the evaluation for these classes.

As stated in a joint paper by HIMSS and SIIM, the implications of our work are multiple (1). An improved method for anatomic labelling of imaging studies has the potential to improve interoperability across healthcare records and systems, address radiology workflow challenges such as labelling discrepancies for studies shared within and between facilities (19), accurate retrieval of anatomically relevant comparison images from image archives and bandwidth related latency and costs associated with data retrieval from cloud-based image archives. Application of our body region labelling could improve display protocols, image synchronization and relevant prior functions in PACS, leading to improved radiologist accuracy and productivity. Beyond image interpretation, it could enable body region dependent AI driven population health initiatives across institutions. When mining the DICOM tags in the test sets for either the "BodyPartExamined" DICOM tag (BP) or "ProcedureType" (PT), the body region information at the study level was only 22.3% (BP) and 42.2% (PT) accurate for CT, and 58.3% (BP) and 47.8% (PT) accurate for MRI. In this cohort, the anatomical AI has the potential to improve the search for anatomically matched cases for more than 50% of the cases.

Automatic identification of body regions in CT and MRI studies in a general population is a challenging task due to the large spectrum of demographics, comorbidities, acquisition protocols, and imaging artifacts. In this context, the anatomical AI demonstrated state-of-the-art performance with an overall image-level classification accuracy greater than 90%, and performance metrics robust across all body regions and confounding factors such as institution, gender, contrast, manufacturer, slice thickness, CT kernel, and MRI sequences. Our next step is to extend the evaluation to other data sources and conduct an observer study to assess the impact on radiologist workflows.

| Body Region | Anatomical landmarks (top) | Anatomical landmarks (bottom) |
|---|---|---|
| Abdomen | Diaphragm / Lung base | Bifurcation of the aorta |
| Breast (MRI only) | Skin surface of upper breast at chest wall | Skin surface of lower breast at chest wall |
| Calf/Lower leg | Proximal $6^{th}$ of tibia | Distal third of calf |
| Chest | Lung apex | Lung base |
| Elbow | Distal 6th of humerus | Proxima 6th of radius |
| Foot | Distal third of calf | Bottom of the foot |
| Forearm | Proximal $6^{th}$ of radius | Distal $6^{th}$ of radius |
| Hand | Distal $6^{th}$ of Radius | Tip of finger |
| Head | Top of head | Bottom of skull base (foramen magnum) |
| Humerus/Arm | Proximal $6^{th}$ of humerus | Distal 6th of humerus |
| Knee | Distal $6^{th}$ of femur | Proximal 6th of tibia |
| Neck | Skull base (foramen magnum) | Lung apex |
| Pelvis & Hip | Aortic bifurcation | Lesser trochanter (hip), inferior extent of pubis symphysis (pelvis) |
| Shoulder | Top of Acromioclavicular (AC) Joint | Proximal 6th of humerus |
| Spine cervical | Tip of odontoid | Bottom of T1 |
| Spine thoracic | Top of T1 | Bottom of T12 |
| Spine lumbar | T11 | Mid sacrum (S2, S3) |
| Thigh/Upper leg | Lesser trochanter | Distal $6^{th}$ of femur |

**Table 1:** Anatomical landmarks for all 18 body region classes.

**Fig 1**: Labeling can be done on any cross-section series (axial, sagittal, coronal, oblique). In this example a rectangle is first drawn on the axial plane to indicate the pelvis area and then extended on the coronal viewport to the lesser trochanter. This creates a thin 3d bonding box which can be easily manipulated in all dimensions to cover the whole anatomy. The 3d bounding box label is automatically carried over to all other series within the same frame of reference using the patient coordinate system. With this technique hundreds of images can be labeled in few seconds with a handful of clicks. If needed, AI Body Region image inference results can be made available with a color code associated with each anatomical class (see bottom color bar). For this image, the AI prediction indicates a pelvis (pink) with a confidence level of 0.825. Lower in the scan, thigh images (purple) are also correctly predicted.

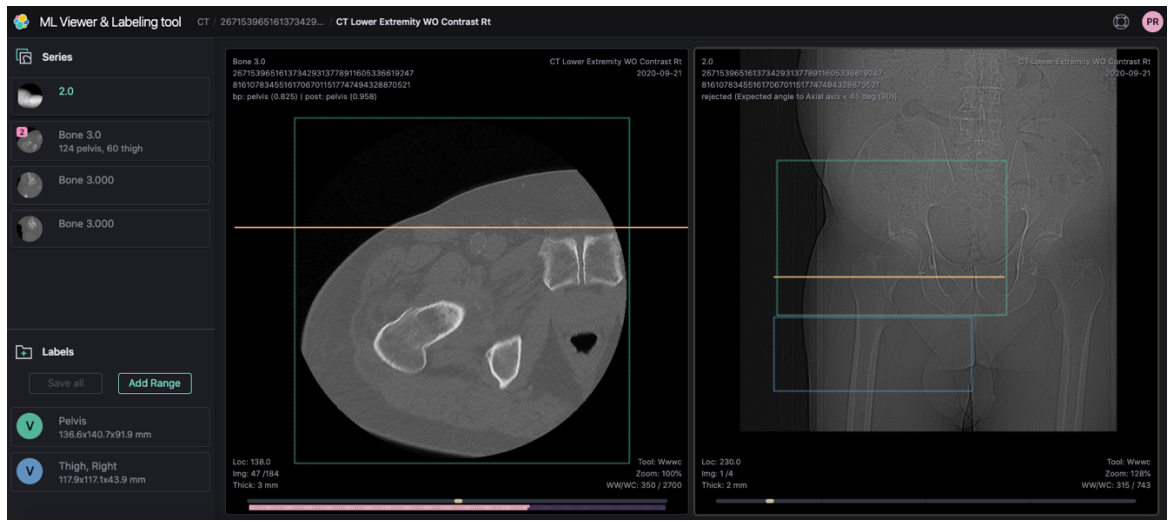

**Fig. 2:** Body region classification steps. In order to accommodate the model's input, several pre-processing steps were applied at the image level. First, pixel values were clipped to fit the interval [mean – 4*std, mean + 4*std], where mean and std correspond to the mean value and standard deviation of pixels in the image. Second, pixel values were normalized using the following transform (pixel value – mean)/(2*std). Third, the gray scale images were converted to RGB images by copying pixel data from first channel to the second and third channels. Last, each image was resized with zero padding to fit the model's required input size of 224 x 224 pixels. Following the 2D CNN image classifier, two rules were applied at the series level to merge the abdomen-chest class to either the abdomen or chest class depending on which body region was predominant. In the absence of chest and abdomen predictions, an abdomen-chest prediction was classified as abdomen. A second rule was classified as breast when at least 50% of the images in the series were classified as such. This is to eliminate spurious measurements in noisy breast acquisitions. In order to discard series with little anatomical content and severe artifacts, a routine was applied to remove MRI series with uncertain results based on the evaluation of entropy. In the last two stages of post-processing, we first applied a routine at the series level to remove outlier labels. We then applied a moving average filter with a window size of three pixels to smooth out results so a continuity in classified labels could be observed throughout the series.

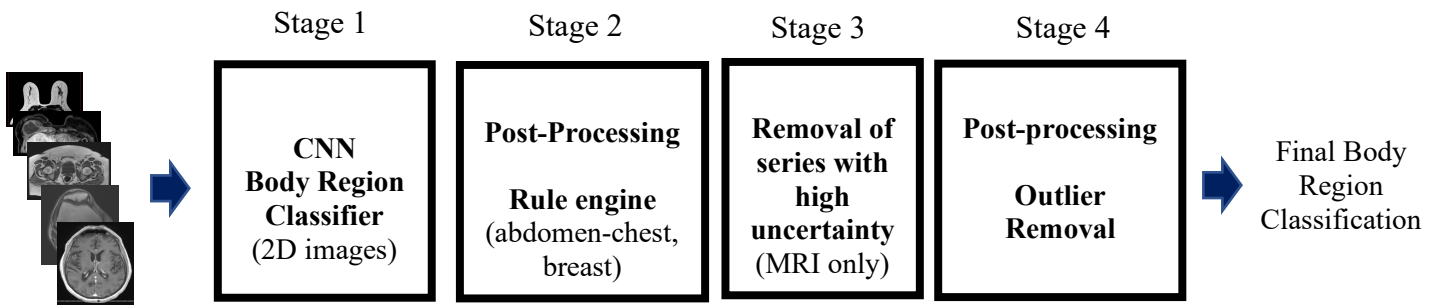

|  | Category | n (%) | Sensitivity (95% CI) | Specificity (95% CI) | p-value |
|---|---|---|---|---|---|
| **Institution** | Primary Care Hospital | 272 (56.1) | 92.1 (91.7 – 92.5) | 99.5 (99.5 - 99.5) | 0.119 |
|  | Community Hospital | 111 (22.9) | 91.5 (90.8 - 92.2) | 99.4 (99.3 - 99.4) |  |
|  | Imaging center | 102 (21.0) | 94.1 (93.6 - 94.6) | 97.1 (96.8 - 97.4) |  |
| **Age** | 18 – 44 years | 103 (21.2) | 91.0 (90.3 - 91.8) | 99.4 (99.4 - 99.5) | 6.7e-7 |
|  | 45 – 64 years | 161 (33.2) | 91.5 (91.0 - 92.1) | 99.4 (99.3 - 99.4) |  |
|  | ≥ 65 years | 221 (45.6) | 93.7 (93.3 - 94.1) | 99.5 (99.5 - 94.5) |  |
| **Gender** | Female | 245 (50.5) | 91.8 (91.4 - 92.3) | 99.4 (99.4 - 99.4) | 0.004 |
|  | Male | 240 (49.5) | 93.1 (92.7 - 93.5) | 99.5 (99.5 - 99.5) |  |
| **Manufacturer** | Canon | 2 (0.4) | NA | NA | 0.003 |
|  | GE | 62 (12.4) | 93.6 (92.7 - 94.4) | 99.5 (99.4 - 99.5) |  |
|  | Hitachi | 4 (0.8) | NA | NA |  |
|  | Philips | 23 (4.6) | 93.0 (91.6 - 94.5) | 99.3 (99.1 - 99.5) |  |
|  | Siemens | 343 (68.5) | 92.3 (92.0 - 92.7) | 99.3 (99.3 - 99.4) |  |
|  | Toshiba | 50 (10.0) | 91.3 (90.2 - 92.4) | 99.4 (99.3 - 99.5) |  |
|  | Vital Images | 17 (3.4) | 100.0 (90.0 - 100.0) | 100.0 (98.1 - 100.0) |  |
| **Contrast*** | With contrast | 305 (26.1) | 94.9 (94.5 - 95.4) | 98.8 (98.7 - 99.0) | 7.0e-8 |
|  | Without contrast | 865 (73.9) | 91.4 (91.1 - 91.8) | 99.5 (99.5 - 99.5) |  |
| **Slice thickness*** | ≤2 mm | 464 (39.7) | 92.7 (92.3 - 93.1) | 99.6 (99.5 - 99.6) | 7.7e-9 |
|  | >2 mm and <5 mm | 641 (54.8) | 91.9 (91.5 - 92.3) | 99.1 (99.1 - 99.2) |  |
|  | ≥5 mm | 64 (5.5) | 96.3 (95.4 - 97.1) | 98.3 (97.6 - 99.0) |  |
| **CT kernel*** | Bone | 268 (22.9) | 92.0 (91.4 - 92.6) | 99.5 (99.4 - 99.5) | 0.623 |
|  | Soft tissue | 901 (77.1) | 92.6 (92.3 - 93.0) | 99.3 (99.3 - 99.4) |  |

**Table 2**: CT image performance metrics by confounding factors. n=number of studies (*series). The p-value for the median chi-square is provided to determine if a significant difference in accuracy is found for each confounding factor.

|  | Category | n (%) | Sensitivity (95% CI) | Specificity (95% CI) | p-value |
|---|---|---|---|---|---|
| **Institution** | Primary Care Hospital | 435 (54.9) | 92.4 (92.1 - 92.8) | 99.2 (99.1 - 99.2) | 9.7e-16 |
|  | Community Hospital | 164 (20.7) | 91.0 (90.4 - 91.5) | 98.5 (98.4 - 98.7) |  |
|  | Imaging center | 69 (8.7) | 93.9 (92.8 - 94.8) | 99.6 (99.6 - 99.7) |  |
|  | Unknown** | 124 (15.7) | 94.4 (93.5 - 95.4) | 99.3 (99.1 - 99.5) |  |
| **Age** | 18 – 44 years | 207 (26.1) | 95.8 (95.3 - 96.2) | 99.6 (99.6 - 99.7) | 2.1e-29 |
|  | 45 – 64 years | 334 (42.2) | 91.4 (90.9 - 91.8) | 99.2 (99.2 - 99.3) |  |
|  | ≥ 65 years | 251 (31.7) | 90.9 (90.5 - 91.4) | 98.7 (98.6 - 98.8) |  |
| **Gender** | Female | 434 (56.1) | 92.0 (91.6 - 92.3) | 99.1 (99.0 - 99.1) | 0.014 |
|  | Male | 339 (43.9) | 92.6 (92.2 - 93.0) | 99.3 (99.2 - 99.3) |  |
| **Manufacturer** | GE | 131 (16.5) | 90.3 (89.7 - 91.0) | 98.1 (97.9 - 98.3) | 6.4e-21 |
|  | Hitachi | 6 (0.8) | NA | NA |  |
|  | Philips | 40 (5.0) | 95.1 (94.3 - 95.8) | 98.7 (98.5 - 99.0) |  |
|  | Siemens | 565 (71.2) | 93.0 (92.7 - 93.3) | 99.4 (99.4 - 99.4) |  |
|  | Toshiba | 50 (6.3) | 93.4 (92.1 - 94.7) | 99.6 (99.4 - 99.7) |  |
| **Contrast*** | With contrast | 343 (16.0) | 91.9 (91.3 - 92.6) | 98.8 (98.7 - 98.9) | 0.115 |
|  | Without contrast | 1,805 (84.0) | 92.3 (92.0 - 92.6) | 99.2 (99.2 - 99.3) |  |
| **Slice thickness*** | ≤2 mm | 114 (5.3) | 98.7 (98.2 - 99.1) | 99.9 (99.9 - 100.0) | 2.4e-27 |
|  | >2 mm and <5 mm | 1,472 (68.5) | 91.9 (91.6 - 92.3) | 99.2 (99.2 - 99.3) |  |
|  | ≥5 mm | 562 (26.2) | 91.5 (91.0 - 92.0) | 98.7 (98.6 - 98.8) |  |
| **MRI Sequence*** | Image weighting | 1,499 (69.8) | 93.1 (92.8 - 93.5) | 99.6 (99.6 - 99.6) | 5.0e-116 |
|  | Spin echo | 63 (2.9) | 92.0 (90.3 - 93.6) | 96.2 (95.2 – 97.2) |  |
|  | Gradient echo | 246 (11.5) | 90.3 (89.7 - 91.0) | 97.0 (96.8 - 97.3) |  |
|  | Inversion recovery | 116 (5.4) | 94.5 (93.3 - 95.7) | 99.6 (99.5 - 99.7) |  |
|  | MRA | 39 (1.8) | 94.8 (92.8 - 96.6) | 97.2 (96.0 - 98.2) |  |
|  | In and Out of Phase | 14 (0.7) | NA | NA |  |
|  | Diffusion | 34 (1.6) | 82.3 (80.2 - 84.3) | 90.1 (88.7 - 91.4) |  |

| | | | | |
|---|---|---|---|---|
| | Unknown*** | 137 (6.4) | 94.1 (93.5 - 94.7) | 97.9 (97.6 - 98.2) | |

Table 3: MRI image performance metrics by confounding factors. n=number of studies (*series). The p-value for the median chi-square is provided to determine if a significant difference in accuracy is found for each confounding factor. **124 studies did not have institution information. ***137 series did not have any of the preset sequence tags. Due to the small number of cases, the performance metrics and confidence interval are not reliable for "In and Out of Phase".

| Body Region | CT | | | MRI | | |
| --- | --- | --- | --- | --- | --- | --- |
| | n | Sensitivity (95% CI) | Specificity (95% CI) | n | Sensitivity (95% CI) | Specificity (95% CI) |
| Overall | 262,326 | 92.5 (92.1 - 92.8) | 99.4 (99.4 - 99.5) | 118,829 | 92.3 (92.0 – 92.5) | 99.2 (99.1 – 99.2) |
| Abdomen | 22,302 | 96.7 (96.1 - 97.3) | 98.7 (98.6 - 98.8) | 17,517 | 92.4 (91.7 - 93.0) | 97.5 (97.3 - 97.6) |
| Arm | 14,430 | 94.1 (92.9 - 95.3) | 99.2 (99.1 - 99.3) | 3,815 | 88.8 (87.1 - 90.5) | 99.8 (99.8 - 99.8) |
| Breast | - | - | - | 20,501 | 100.0 (100.0 - 100.0) | 100.0 (100.0 - 100.0) |
| Calf | 12,308 | 93.0 (91.7 - 94.3) | 99.7 (99.6 - 99.7) | 3,299 | 95.7 (94.5 - 96.9) | 99.6 (99.6 - 99.7) |
| Chest | 27,968 | 96.9 (96.3 - 97.6) | 98.9 (98.8 - 99.0) | 13,010 | 89.2 (88.4 - 90.1) | 98.5 (98.4 - 98.7) |
| Cervical spine | 11,993 | 78.0 (76.0 - 80.0) | 99.9 (99.9 - 99.9) | 3,607 | 62.3 (59.3 - 65.5) | 99.9 (99.9 - 99.9) |
| Elbow | 11,288 | 88.0 (86.0 - 90.0) | 99.6 (99.6 - 99.7) | 4,615 | 91.2 (89.6 - 92.7) | 99.7 (99.6 – 99.7) |
| Foot | 13,137 | 86.5 (84.8 - 88.2) | 99.9 (99.8 - 99.9) | 3,178 | 94.8 (93.3 - 96.4) | 99.9 (99.9 - 99.9) |
| Forearm | 8,232 | 86.6 (84.3 - 88.8) | 99.5 (99.5 - 99.6) | 3,546 | 86.0 (84.1 - 87.9) | 99.7 (99.7 - 99.8) |
| Hand | 11,321 | 94.9 (93.7 - 96.2) | 99.6 (99.5 - 99.6) | 6,699 | 96.3 (95.4 - 97.3) | 99.8 (99.7 - 99.8) |
| Head | 29,066 | 98.7 (98.3 - 99.1) | 99.9 (99.9 - 99.9) | 7,633 | 99.3 (98.8 - 99.7) | 99.7 (99.7 - 99.8) |
| Knee | 10,721 | 94.0 (92.7 - 95.3) | 99.4 (99.3 - 99.4) | 3,649 | 95.8 (94.5 - 97.0) | 99.8 (99.8 - 99.9) |
| Lumbar spine | 16,100 | 93.6 (92.4 - 94.8) | 99.8 (99.8 - 99.9) | 4,556 | 97.0 (96.0 - 98.0) | 98.8 (99.8 - 99.9) |
| Neck | 19,292 | 94.4 (93.3 - 95.5) | 98.9 (98.8 - 99.0) | 4,363 | 86.1 (83.7 - 88.6) | 98.9 (98.8 - 99.0) |
| Pelvis | 13,886 | 84.9 (83.2 - 86.6) | 99.8 (99.7 - 99.8) | 11,632 | 90.3 (89.4 - 91.3) | 99.4 (99.3 - 99.5) |
| Shoulder | 13,266 | 88.6 (86.9 - 90.2) | 99.8 (99.7 - 99.8) | 5,599 | 98.6 (98.0 - 99.2) | 99.6 (99.6 - 99.7) |
| Thigh | 14,409 | 89.4 (88.0 - 90.8) | 99.9 (99.9 - 99.9) | 4,485 | 88.9 (87.3 - 90.5) | 99.9 (99.9 - 100.0) |
| Thoracic spine | 21,378 | 91.4 (90.3 - 92.5) | 99.5 (99.5 - 99.6) | 3,952 | 93.4 (91.9 - 94.8) | 99.7 (99.7 - 99.8) |

**Table 4:** CT and MRI image classification sensitivity and specificity by body region. n=number of images.

# SUPPLEMENTAL MATERIALS

## Inclusion and Exclusion Criteria

Many types of MRI sequences were included in the study such as image weighting, spin echo, gradient echo, inversion recovery, MRA, In and Out of Phase, and diffusion imaging (Table 1).

Were excluded from consideration in the study, imaging sequences used to determine flow velocities such as VIPR (High Speed 3D Phase-Contrast method for flow) and FASTPC (Fast Phase Contrast Imaging) as well as the dynamic MRA sequence called TRICKS (Time Resolved Imaging of Contrast KineticS) and MRE (MR Elastography) used for liver disease. Regarding CT, only series related to the evaluation of flow velocities were excluded. For both modalities, a first filtering pass was applied using a search on selective keywords in the series description (Table 2). A second quality control review was conducted to ensure no irrelevant series were kept in the training database.

| MRI Sequence Type | Sequence Names |
|---|---|
| Image weighting | T1, T2, T1/T2*, PD, SWI, SWAN, BRAVO, MERGE |
| Spin echo | SE, Fast SE, Single Shot FSE, HASTE, VISTA, PROPELLER |
| Gradient echo | GRE, FFE, FE, SPGR, FGRE, LAVA, VIBRANT, VIBE, BLISS, FISP, SSFP, FIESTA, TRU FISP |
| Inversion recovery | IR, STIR, FLAIR |
| MRA | TOF, CONTRAST ENHANCED, DELAY ENHANCED |
| In and Out of Phase | IN/OUT, IN of Phase, Out of Phase |
| Diffusion | DIFFUSION, DWI |

**Table 1**: Examples of MRI sequences included in the development databases

|  | **EXCLUDED SEQUENCES WITH FOLLOWING KEYWORDS** |
| --- | --- |
| **MRI series** | FLOW, VELOCITY, ADC, APPARENT DIFFUSION, IDEAL, EP2D_DIFF, FASTPC, PC, CBV, CBF, MTT, TTP, CAD, DWI_SSH, VIPR |
| **CT series** | VELOCITY |

**Table 2**: Series filtering using selective keywords in the series description

The flowcharts of exclusion criteria for both modalities can be found in the following figures below:

- Figure 1: CT exclusion criteria (development databases)
- Figure 2: MRI exclusion criteria (development databases)
- Figure 3: CT exclusion criteria (Test database)
- Figure 4: MRI exclusion criteria (Test database)

**Fig 1**: Flowchart of exclusion criteria for CT annotated images (development databases).

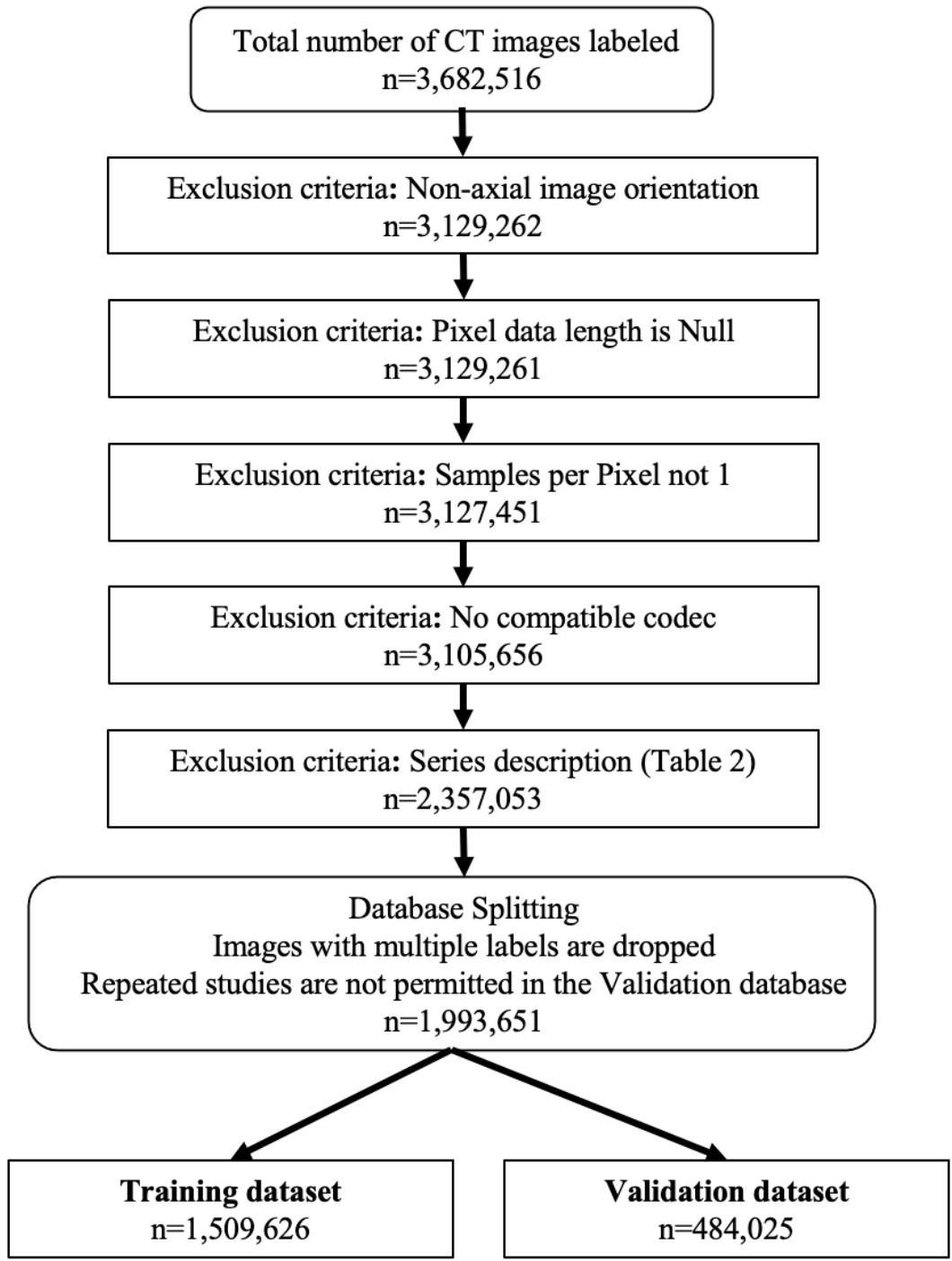

**Fig 2**: Flowchart of exclusion criteria for MRI annotated images (development databases).

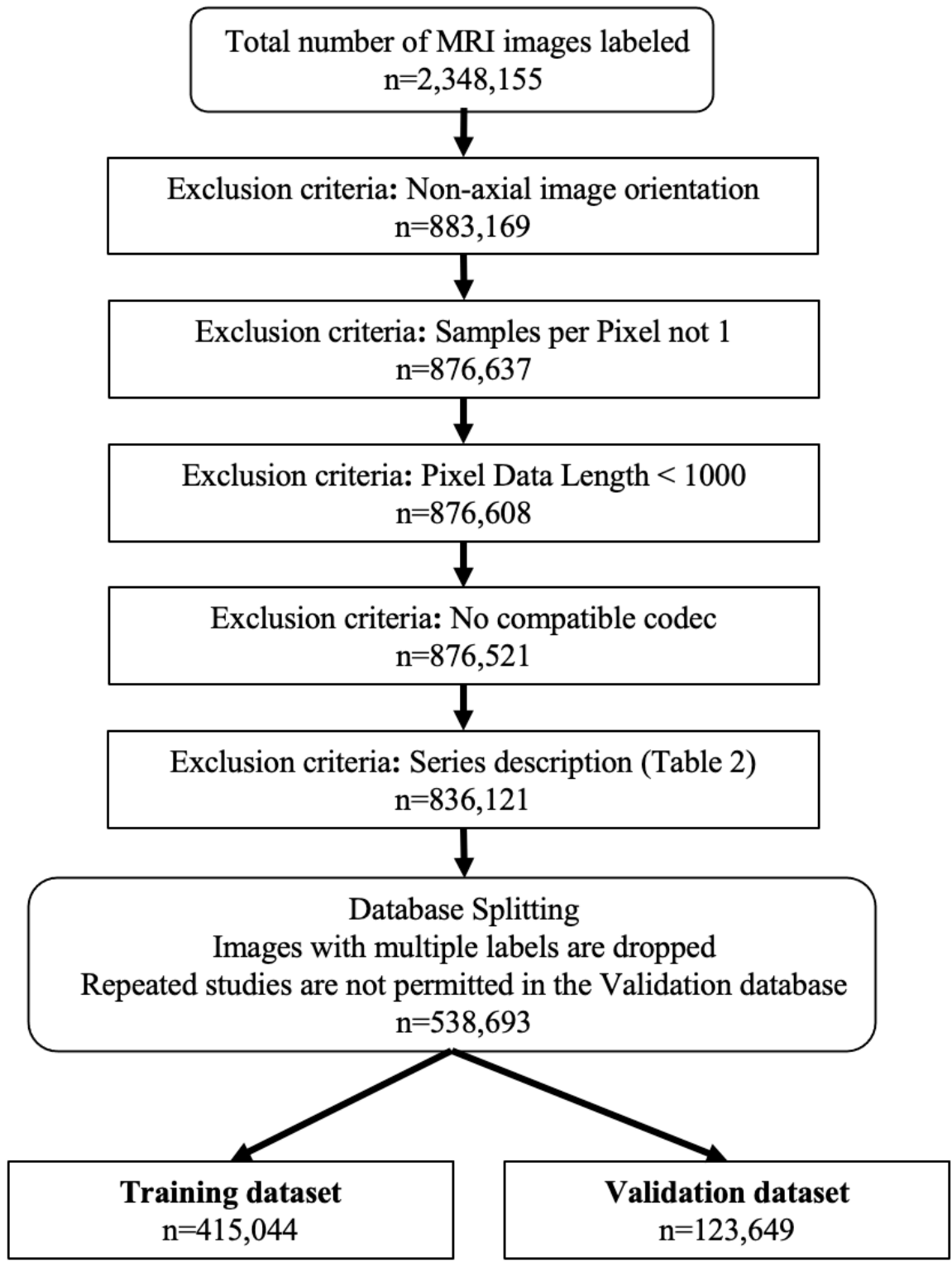

**Fig 3**: Flowchart of exclusion criteria for CT annotated images (Test database).

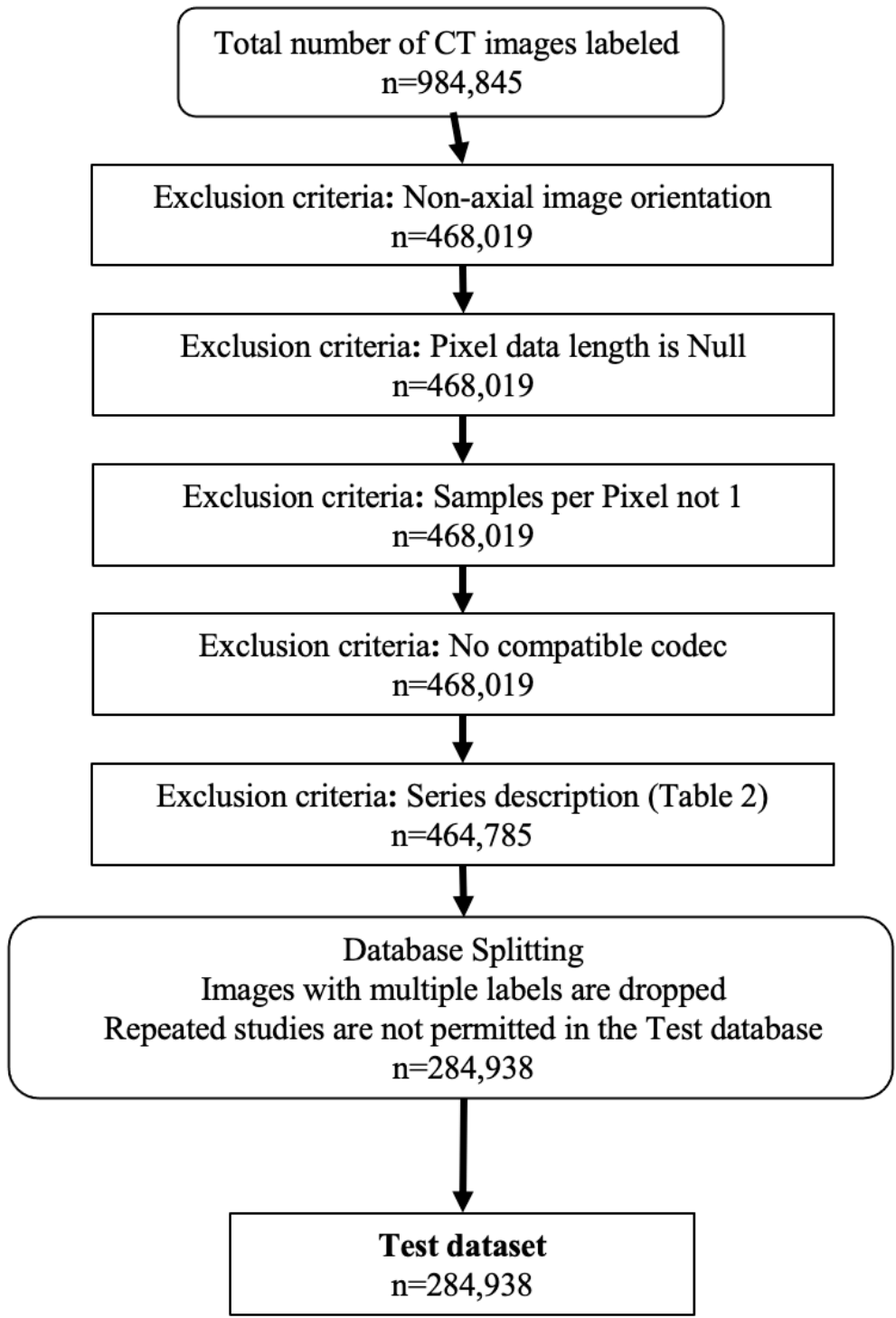

**Fig 4**: Flowchart of exclusion criteria for MRI test annotated images (Test database).

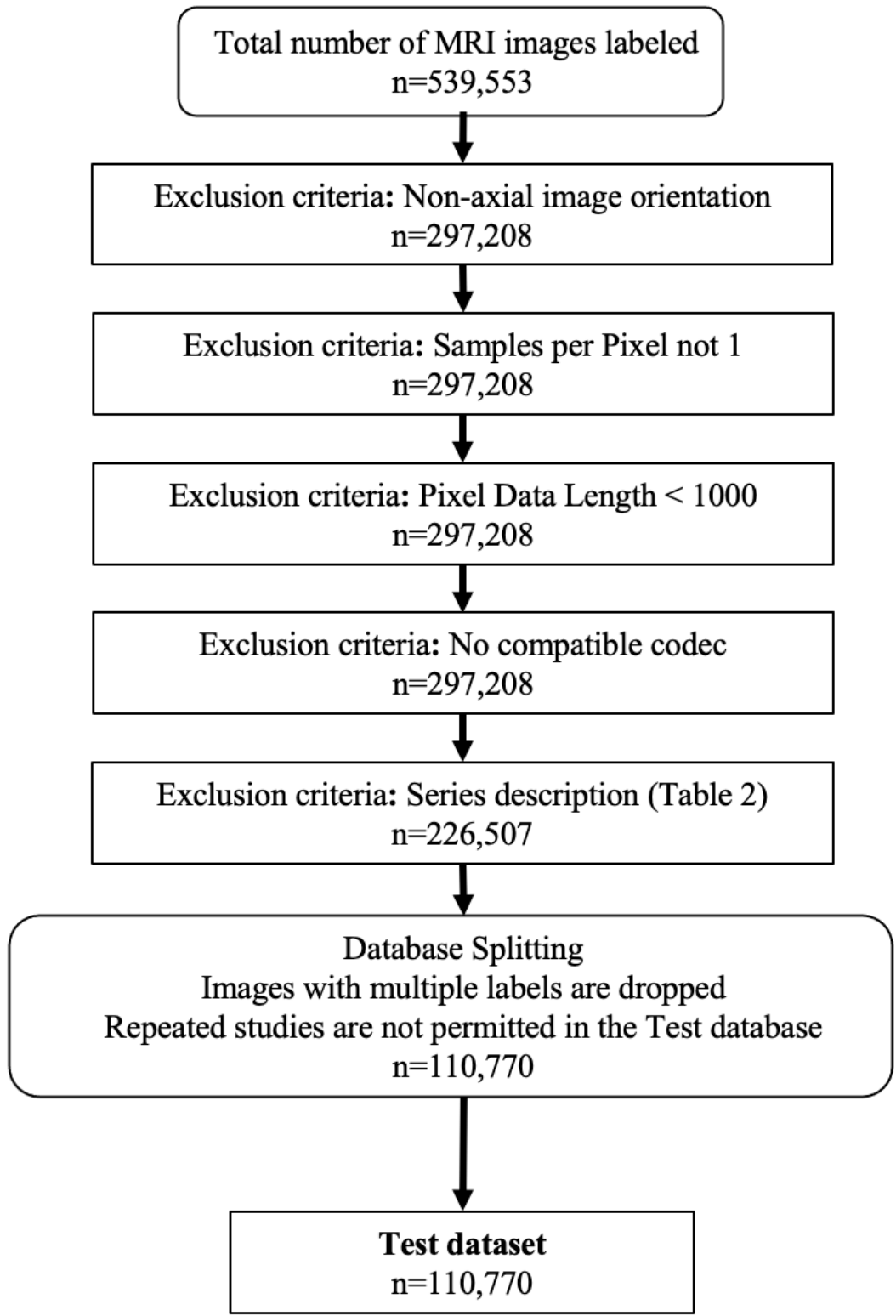

## Training Parameters

| Parameters | CT Model | MRI model |
|---|---|---|
| Best Training epoch | 87 | 89 |
| Learning rate | 1e-05 | 1e-05 |
| Optimization algorithm | Nadam | Nadam |
| Batch size | 128 | 96 |
| Step size | 2948 | 865 |

**Table 3**: Training hyperparameters

## Distribution Development Databases

- Table 4: patient and image distribution by body region (CT)
- Table 5: patient and image distribution by body region (MRI)
- Table 6: study distribution by institution, demographics, and manufacturer (CT)
- Table 7: series distribution by scanner parameters and clinical protocol (CT)
- Table 8: study distribution by institution, demographics, and manufacturer (MRI)
- Table 9: series distribution by scanner parameters and clinical protocol (MRI)

| Body Region | Patients | | Images | |
| --- | --- | --- | --- | --- |
| | Training | Validation | Training | Validation |
| Head | 262 | 86 | 78,487 | 20,732 |
| Neck | 315 | 107 | 54,691 | 17,483 |
| Cervical spine | 209 | 68 | 52,470 | 17,467 |
| Chest | 510 | 165 | 105,262 | 34,745 |
| Thoracic spine | 285 | 95 | 74,552 | 24,566 |
| Abdmchst | 356 | 117 | 48,660 | 17,854 |
| Abdomen | 354 | 113 | 58,808 | 19,154 |
| Lumbar spine | 147 | 49 | 65,147 | 21,098 |
| Pelvis | 319 | 96 | 62,556 | 19,018 |
| Thigh | 228 | 69 | 88,390 | 25,663 |
| Knee | 227 | 70 | 131,590 | 43,054 |
| Calf | 262 | 82 | 150,307 | 46,935 |
| Foot | 245 | 81 | 149,354 | 47,647 |
| Shoulder | 290 | 102 | 70,008 | 24,832 |
| Arm | 361 | 126 | 72,357 | 23,209 |
| Elbow | 237 | 82 | 77,044 | 25,847 |
| Forearm | 254 | 90 | 49,548 | 17,515 |
| Hand | 250 | 79 | 120,395 | 37,206 |

**Table 4**: Distribution of CT patients as a function of body region information extracted from the ground-truth. Note that a study can contain several body regions, as a result it may contribute to several anatomical regions.

| Body Region | Patients | | Images | |
|---|---|---|---|---|
| | Training | Validation | Training | Validation |
| Head | 208 | 66 | 60,027 | 17,474 |
| Neck | 249 | 84 | 15,838 | 4,923 |
| Cervical spine | 135 | 43 | 5,976 | 1,989 |
| Chest | 330 | 108 | 17,995 | 4,895 |
| Thoracic spine | 185 | 63 | 98,27 | 3,226 |
| Abdmchst | 265 | 89 | 18,672 | 5,649 |
| Abdomen | 253 | 78 | 40,835 | 11,680 |
| Lumbar spine | 134 | 45 | 7,660 | 2,094 |
| Pelvis | 288 | 96 | 31,777 | 8,783 |
| Thigh | 201 | 55 | 11,795 | 3,272 |
| Knee | 219 | 74 | 7,898 | 2,425 |
| Calf | 138 | 44 | 11,160 | 3,701 |
| Foot | 224 | 75 | 10,433 | 3,082 |
| Shoulder | 176 | 62 | 6,380 | 1,945 |
| Arm | 180 | 61 | 7,844 | 2,432 |
| Elbow | 325 | 100 | 16,926 | 4,911 |
| Forearm | 225 | 72 | 11,342 | 3,778 |
| Hand | 223 | 69 | 14,263 | 4,234 |
| Breast | 137 | 46 | 108,396 | 33,156 |

**Table 5**: Distribution of MRI patients as a function of body region information extracted from the ground-truth. Note that a study can contain several body regions, as a result it may contribute to several anatomical regions.

| Attributes | Categories | Training | | Validation | |
|---|---|---|---|---|---|
| | | n | % | n | % |
| **Gender** | Female | 867 | 48.1 | 294 | 48.8 |
| | Male | 935 | 51.9 | 308 | 51.2 |
| **Age** | 18 - 44 years | 436 | 24.2 | 133 | 22.1 |
| | 45 – 64 years | 640 | 35.5 | 236 | 39.3 |
| | ≥ 65 years | 654 | 36.3 | 212 | 35.3 |
| **Manufacturer** | Canon | - | - | - | - |
| | GE | 1,671 | 92.5 | 555 | 92.3 |
| | Hitachi | - | - | - | - |
| | Philips | 4 | 0.2 | 2 | 0.3 |
| | Siemens | 96 | 5.3 | 37 | 6.2 |
| | Toshiba | 35 | 1.9 | 7 | 1.2 |
| | Others | - | - | - | - |

**Table 6**: Distribution of CT studies with respect to demographics, and manufacturer.

| Attributes | Categories | Training | | Validation | |
|---|---|---|---|---|---|
| | | n | % | n | % |
| Contrast | With contrast | 1,300 | 32.6 | 404 | 31.1 |
| | Without contrast | 2689 | 67.4 | 894 | 68.9 |
| Slice thickness | ≤ 2 mm | 2,722 | 67.2 | 899 | 68.1 |
| | >2 mm and < 5 mm | 343 | 8.5 | 98 | 7.4 |
| | ≥ 5 mm | 985 | 24.3 | 324 | 24.5 |
| Kernel | Bone | 1,072 | 26.9 | 342 | 26.3 |
| | Soft tissue | 2,917 | 73.1 | 956 | 73.7 |

**Table 7**: Distribution of CT series with respect to clinical protocol. Series with a bone CT kernel were identified using a search of the keyword "bone" in the series description. It includes series with and without Metal Artifact Reduction (MAR). All other axial series in the same study were considered soft tissue acquisitions.

| Attributes | Categories | Training | | Validation | |
|---|---|---|---|---|---|
| | | n | % | n | % |
| Gender | Female | 1,040 | 54.5 | 326 | 52.2 |
| | Male | 870 | 45.5 | 298 | 47.8 |
| Age | 18 – 44 years | 592 | 31.6 | 194 | 31.5 |
| | 45 – 64 years | 668 | 35.6 | 214 | 34.3 |
| | ≥ 65 years | 375 | 20 | 130 | 21.1 |
| Manufacturer | GE | 1,634 | 85.5 | 542 | 86.7 |
| | Hitachi | - | - | 2 | 0.3 |
| | Philips | 40 | 2.1 | 12 | 1.9 |
| | Siemens | 235 | 12.3 | 69 | 11 |
| | Toshiba | 1 | 0.1 | - | - |

**Table 8**: Distribution of MRI studies with respect to demographics, and manufacturer.

| Attributes | Categories | Training | | Validation | |
|---|---|---|---|---|---|
| | | n | % | n | % |
| Contrast | With contrast | 1585 | 22.4 | 520 | 24.1 |
| | Without contrast | 5466 | 77.6 | 1641 | 75.9 |
| Slice thickness | ≤ 2 mm | 934 | 13.2 | 287 | 13.3 |
| | >2 mm and < 5 mm | 3,402 | 48.2 | 1,102 | 51.0 |
| | ≥ 5 mm | 2,715 | 38.5 | 772 | 35.7 |
| Sequences | Image weighting | 5,378 | 73.2 | 1,656 | 73.9 |
| | Spin echo | 329 | 4.5 | 127 | 5.7 |
| | Gradient echo | 821 | 11.2 | 234 | 10.4 |
| | Inversion recovery | 221 | 3.0 | 57 | 2.5 |
| | MRA | 182 | 2.5 | 40 | 1.8 |
| | In and Out of Phase | 108 | 1.5 | 38 | 1.7 |
| | Diffusion | 307 | 4.2 | 88 | 3.9 |

**Table 9**: Distribution of MRI series with respect to clinical protocol. MRI sequences are selected based on the search of keywords in series description listed in Table 1.

# Results Validation Database

- Table 10: sensitivity and specificity by body region (CT)
- Table 11: sensitivity and specificity by body region (MRI)
- Table 12: sensitivity and specificity by institution, demographics, manufacturer, scanner parameters (CT)
- Table 13: sensitivity and specificity by institution, demographics, manufacturer, scanner parameters (MRI)

| Body Region | Sensitivity (95% CI) | Specificity (95% CI) |
|---|---|---|
| Overall | 91.7 (91.5 – 91.9) | 99.3 (99.3 – 99.4) |
| Head | 98.7 (98.3 – 99.1) | 99.9 (99.8 – 99.9) |
| Neck | 89.4 (88.1 – 90.7) | 99.8 (99.7 – 99.8) |
| Cervical spine | 91.6 (90.1 – 93.1) | 99.8 (99.7 – 99.8) |
| Chest | 92.6 (92.0 – 93.1) | 98.9 (98.8 – 98.9) |
| Thoracic spine | 95.6 (94.7 – 96.6) | 99.6 (99.5 – 99.6) |
| Abdomen | 93.3 (92.8 – 93.9) | 98.8 (98.7 – 98.9) |
| Lumbar spine | 96.4 (95.5 – 97.2) | 99.8 (99.8 – 99.9) |
| Pelvis | 90.5 (89.4 – 91.5) | 99.5 (99.5 – 99.6) |
| Thigh | 88.2 (86.9 – 89.4) | 99.3 (99.2 – 99.4) |
| Knee | 82.1 (80.7 – 83.4) | 99.5 (99.4 – 99.5) |
| Calf | 91.2 (90.2 – 92.2) | 99.3 (99.2 – 99.3) |
| Foot | 94.2 (93.4 – 95.1) | 99.6 (99.6 – 99.7) |
| Shoulder | 91.4 (90.3 – 92.3) | 99.5 (99.4 – 99.5) |
| Arm | 90.4 (89.2 – 91.5) | 99.1 (99.0 – 99.2) |
| Elbow | 85.6 (84.1 – 87.2) | 99.6 (99.5 – 99.6) |
| Forearm | 87.2 (85.6 – 88.8) | 99.4 (99.3 – 99.5) |
| Hand | 92.4 (91.5 – 93.4) | 99.8 (99.8 – 99.8) |

**Table 10**: Validation database - CT weighted image-level sensitivity and specificity (%) as a function of body region.

| Body Region | Sensitivity (95% CI) | Specificity (95% CI) |
|---|---|---|
| Overall | 93.9 (93.7 – 94.3) | 99.3 (99.3 – 99.3) |
| Head | 98.9 (98.4 – 99.4) | 99.6 (99.5 – 99.7) |
| Neck | 85.4 (82.7 – 88.0) | 99.8 (99.8 – 99.9) |
| Cervical spine | 84.8 (81.9 – 87.7) | 99.9 (99.9 – 99.9) |
| Chest | 89.6 (88.7 – 90.6) | 98.5 (98.3 – 98.6) |
| Thoracic spine | 97.9 (97.0 – 98.8) | 99.7 (99.6 – 99.8) |
| Abdomen | 92.7 (92.2 – 93.6) | 98.0 (97.9 – 98.2) |
| Lumbar spine | 89.8 (87.3 – 92.0) | 99.8 (99.8 – 99.9) |
| Pelvis | 92.1 (91.0 – 93.2) | 99.7 (99.6 – 99.7) |
| Thigh | 91.7 (90.0 – 93.4) | 99.9 (99.9 – 99.9) |
| Knee | 94.4 (92.6 – 96.0) | 99.7 (99.7 – 99.8) |
| Calf | 90.8 (89.1 – 92.7) | 99.8 (99.7 – 99.8) |
| Foot | 94.9 (93.2 – 96.6) | 99.8 (99.8 – 99.9) |
| Shoulder | 97.1 (95.6 – 98.5) | 99.9 (99.9 – 99.9) |
| Arm | 90.2 (87.9 – 92.3) | 99.8 (99.7 – 99.8) |
| Elbow | 92.3 (90.5 – 94.0) | 99.7 (99.6 – 99.8) |
| Forearm | 91.8 (89.8 – 93.8) | 99.8 (99.8 – 99.9) |
| Hand | 97.5 (96.3 – 98.5) | 99.8 (99.8 – 99.9) |
| Breast | 99.9 (99.8 – 99.9) | 99.9 (99.9 – 99.9) |

**Table 11**: Validation database - MRI weighted image-level sensitivity and specificity (%) as a function of body region.

| Attributes | Categories | Sensitivity (95% CI) | Specificity (95% CI) |
|---|---|---|---|
| **Gender** | Female | 91.8 (91.4 – 92.1) | 99.4 (99.3 – 99.4) |
| | Male | 91.7 (91.4 – 92.0) | 99.3 (99.3 – 99.3) |
| **Age** | 18 – 44 years | 90.8 (90.3 – 91.3) | 99.3 (99.3 – 99.4) |
| | 45 – 64 years | 92.2 (91.9 – 92.5) | 99.3 (99.2 – 99.3) |
| | ≥ 65 years | 91.6 (91.1 – 91.9) | 99.4 (99.3 – 99.4) |
| **Manufacturer** | GE | 91.9 (91.7 – 92.2) | 99.3 (99.3 – 99.4) |
| | Philips | - | - |
| | Siemens | 89.4 (88.1 – 90.7) | 98.8 (98.6 – 98.9) |
| | Toshiba | 78.9 (75.3 – 82.4) | 97.9 (97.1 – 98.7) |
| **Contrast** | With contrast | 92.9 (92.6 – 93.3) | 99.0 (98.9 – 99.1) |
| | Without contrast | 90.9 (90.7 – 91.3) | 99.4 (99.3 – 99.4) |
| **Kernel** | Bone | 91.3 (90.8 – 91.7) | 99.4 (99.4 – 99.4) |
| | Soft tissue | 91.9 (91.6 – 92.1) | 99.1 (99.1 – 99.2) |
| **Slice thickness** | ≤ 2 mm | 91.3 (90.9 – 91.5) | 99.4 (99.4 – 99.4) |
| | > 2 mm and < 5 mm | 91.7 (91.0 – 92.4) | 99.2 (99.1 – 99.3) |
| | ≥ 5 mm | 93.4 (92.9 – 93.8) | 96.9 (96.6 – 97.2) |

**Table 12**: Validation database - CT weighted image-level sensitivity and specificity (%) as a function of demographics, manufacturer and acquisition protocol.

| Attributes | Categories | Sensitivity (95% CI) | Specificity (95% CI) |
|---|---|---|---|
| **Gender** | Female | 94.9 (94.5 – 95.2) | 99.4 (99.4 – 99.5) |
| | Male | 92.3 (92.4 – 93.3) | 99.0 (98.9 – 99.1) |
| **Age** | 18 – 44 years | 94.9 (94.4 – 95.4) | 99.5 (99.5 – 99.6) |
| | 45 – 64 years | 93.9 (93.5 – 94.4) | 99.3 (99.3 – 99.4) |
| | ≥ 65 years | 93.0 (92.5 – 93.5) | 98.9 (98.8 – 99.0) |
| **Manufacturer** | GE | 93.9 (93.6 – 94.2) | 99.4 (99.4 – 99.4) |
| | Hitachi | - | - |
| | Philips | - | - |
| | Siemens | 94.1 (93.4 – 94.7) | 98.3 (98.0 – 98.5) |
| | Toshiba | - | - |
| **Contrast** | With contrast | 94.8 (94.4 – 95.3) | 99.2 (99.1 – 99.3) |
| | Without contrast | 93.5 (93.2 – 93.9) | 99.3 (99.3 – 99.4) |
| **Slice thickness** | ≤ 2 mm | 98.4 (97.9 – 98.8) | 98.9 (98.4 – 99.3) |
| | > 2 mm and < 5 mm | 94.3 (93.6 – 94.5) | 99.5 (99.4 – 99.5) |
| | ≥ 5 mm | 92.5 (92.1 – 92.9) | 98.8 (98.6 – 98.8) |
| **Sequences** | Image weighting | 93.5 (93.1 – 93.9) | 99.5 (99.4 – 99.5) |
| | Spin echo | 91.5 (89.8 – 93.2) | 99.1 (98.9 – 99.3) |
| | Gradient echo | 95.1 (94.5 – 95.6) | 98.6 (98.5 – 98.8) |
| | Inversion recovery | 89.6 (86.9 – 92.3) | 99.3 (99.1 – 99.5) |
| | MRA | 95.9 (93.7 – 97.8) | 97.6 (95.7 – 99.1) |
| | In and Out of Phase | 91.1 (89.8 – 92.4) | 88.9 (87.1 – 90.9) |
| | Diffusion | 95.3 (94.3 – 96.3) | 99.0 (98.7 – 99.3) |

**Table 13**: Validation database - MRI weighted image-level sensitivity and specificity (%) as a function of demographics, manufacturer, and acquisition protocol.